\documentclass{iopjournal}

\usepackage{amsmath,amssymb,mathtools}
\usepackage{siunitx}
\usepackage{hyperref}
\usepackage{booktabs}
\usepackage{graphicx}
\usepackage{subcaption}
\usepackage{float}
\usepackage{array}
\usepackage{tabularx,booktabs}
\newcolumntype{C}{>{\centering\arraybackslash}X}
\usepackage{microtype}
\usepackage{pifont}
\usepackage{hyphenat}
\usepackage{algorithm}
\usepackage{algorithmic}
\usepackage{graphicx}   
\usepackage{booktabs}
\usepackage{float}      
\usepackage{booktabs}
\usepackage{tabularx}
\usepackage{makecell}
\usepackage{array}
\newcolumntype{Y}{>{\centering\arraybackslash}X}

\usepackage{setspace}   
\usepackage{etoolbox}   

\usepackage{enumitem}
\setlist{nosep,leftmargin=*,labelsep=0.5em}


\newcolumntype{L}{>{$}l<{$}} 

\usepackage{ragged2e}                  
\AtBeginDocument{\justifying}          
\begin{document}
\onehalfspacing

\AtBeginEnvironment{figure}{\singlespacing}
\AtBeginEnvironment{table}{\singlespacing}
\AtBeginEnvironment{algorithm}{\singlespacing}
\AtBeginEnvironment{tabular}{\singlespacing}
\AtBeginEnvironment{thebibliography}{\singlespacing}


\title{Privacy-Aware Federated nnU-Net for ECG Page Digitization}

\author{Nader Nemat$^{1,*}$\orcid{0000-0003-2548-6740}}

\affil{$^1$IEEE Machine Learning Member, Turku, Finland}


\affil{$^*$Author to whom any correspondence should be addressed.}

\email{naderr.nemati@outlook.com}


\begin{abstract}
\justifying
Deep neural networks can convert ECG page images into analyzable waveforms, yet centralized training often conflicts with cross-institutional privacy and deployment constraints. A cross-silo federated digitization framework is presented that trains a full-model nnU-Net segmentation backbone without sharing images and aggregates updates across sites under realistic non-IID heterogeneity (layout, grid style, scanner profile, noise).

The protocol integrates three standard server-side aggregators—FedAvg, FedProx, and FedAdam—and couples secure aggregation with central, user-level differential privacy to align utility with formal guarantees. Key features include: (i) end-to-end full-model training and synchronization across clients; (ii) secure aggregation so the server only observes a clipped, weighted sum once a participation threshold is met; (iii) central Gaussian DP with Rényi accounting applied post-aggregation for auditable user-level privacy; and (iv) a calibration-aware digitization pipeline comprising page normalization, trace segmentation, grid-leakage suppression, and vectorization to twelve-lead signals.

Experiments on ECG pages rendered from PTB-XL show consistently faster convergence and higher late-round plateaus with adaptive server updates (FedAdam) relative to FedAvg and FedProx, while approaching centralized performance. The privacy mechanism maintains competitive accuracy while preventing exposure of raw images or per-client updates, yielding deployable, auditable guarantees suitable for multi-institution settings.
\end{abstract}

\section{Introduction}

Electrocardiogram (ECG) page images remain pervasive in clinical archives and day-to-day workflows, while most analytical pipelines assume access to digitally sampled waveforms. Public corpora such as PTB-XL demonstrate the scientific and translational value of curated digital signals for benchmarking and model development. Nevertheless, many health systems retain decades of paper or scanned ECGs that are costly to query and prone to loss \cite{wagner2020ptbxl}. Converting page images into calibrated waveforms preserves longitudinal clinical history, enables secondary analyses at scale, and supports interoperable storage and retrieval across institutions. Recent work has shown that deep learning can recover high-fidelity traces from printed pages by segmenting the trace, suppressing grid artifacts, and reconstructing lead-wise signals with strong agreement to digital ground truth \cite{Wu2022Digitisation,Demolder2024MedRxiv}. The 2024 George B.\ Moody PhysioNet Challenge further catalyzed progress on image-to-signal reconstruction and image-based classification, consolidating best practices for layout-aware post-processing and calibration-aware vectorization \cite{Reyna2024Challenge,Antoni2024CinC,SignalSavants2024ArXiv}.

Centralized training remains the norm for these pipelines but is often infeasible across institutions due to regulatory, governance, and operational constraints. Federated learning (FL) offers a natural alternative by training across sites without moving raw images, with canonical aggregators such as Federated Averaging (FedAvg), proximal regularization (FedProx), and adaptive server optimization in the FedOpt family (FedAdam), addressing non-IID data and client heterogeneity \cite{McMahan2017FedAvg,Li2020FedProx,Reddi2021FedOpt}. In medical imaging and cardiology, cross-site FL has approached centralized performance while preserving data locality, motivating its use for ECG digitization \cite{Sheller2020SciRep}. \textit{More broadly, deep learning on sequential data has proven effective outside biomedicine as well; for example, combining CNNs trained at multiple temporal resolutions improves forecasting performance on financial time series \cite{Nemati2023HORA}.} Yet, evidence specific to \emph{image-to-signal} ECG digitization under federated constraints remains limited compared with time-series classification and echocardiography modeling, leaving open questions about optimization dynamics, utility–privacy trade-offs, and communication efficiency in this setting \cite{Nemati2025RG}.

We adopt a privacy-preserving cross-silo FL design that matches clinical requirements and our optimization pipeline. At each round, clients compute full-model updates on local pages and participate in \emph{secure aggregation (SecAgg)} so that the server can only recover a masked \emph{sum} of clipped updates once a participation threshold is met—individual updates remain hidden via pairwise one-time masks that cancel in aggregate \cite{Bonawitz2017SecAgg}. On the server, we enforce \emph{central} user-level DP by adding Gaussian noise to the aggregated, clipped update and composing privacy loss across rounds with a Rényi moments accountant \cite{Abadi2016DPSGD,Mironov2017RDP}. Compared with local DP that perturbs each client’s update, central DP applied after summation achieves better utility at a fixed privacy target because effective per-client noise scales down with the cohort size; SecAgg alone, while concealing single-site updates, does not bound inference from model histories, hence the combination of SecAgg and central DP, achieves auditable guarantees on the released sequence of aggregates. This mechanism is a drop-in at aggregation time and requires no changes to client-side learning beyond norm clipping, making it compatible with full-model nnU-Net training and standard aggregators, FedAvg, FedProx, and FedAdam, used in this work \cite{McMahan2017FedAvg,Li2020FedProx,Reddi2021FedOpt}.

\section{Related works}

ECG image digitization has progressed from rule-based extraction to fully learned, segmentation-to-vectorization pipelines with high agreement to digital ground truth \cite{Wu2022Digitisation}. Recent systems pair robust page normalization and thin-structure segmentation with calibration-aware vectorization, and community efforts around the 2024 George B.\ Moody PhysioNet Challenge focused the task and released stronger baselines and artifacts for benchmarking \cite{Demolder2024MedRxiv,Reyna2024Challenge,Antoni2024CinC,SignalSavants2024ArXiv}. Within these pipelines, nnU\hbox{-}Net is a frequent backbone due to its self-configuration and strong biomedical-segmentation performance \cite{Isensee2021nnUNet}. Challenge materials and contemporaneous datasets emphasize realistic renderings, scanner artifacts, and layout variability to stress-test reconstruction, consistent with broader medical-imaging work that reports strong centralized baselines but growing interest in distributed training when governance constrains data pooling \cite{Sheller2020SciRep}.

Federated learning (FL) offers an alternative to centralization under cross-site heterogeneity. Canonical optimizers include sample-size–weighted Federated Averaging (FedAvg), proximal regularization with FedProx, and adaptive server methods in the FedOpt family (FedAdam), each mitigating client drift to different degrees \cite{McMahan2017FedAvg,Li2020FedProx,Reddi2021FedOpt}. To address privacy, secure aggregation (SecAgg) prevents the server from inspecting any single client update by revealing only a masked sum after a participation threshold \cite{Bonawitz2017SecAgg}. Because SecAgg alone does not bound inferences from the released aggregates or model history, many deployments combine it with central DP, adding calibrated Gaussian noise to the post-aggregation vector and composing privacy loss across rounds via Rényi accounting \cite{Abadi2016DPSGD,Mironov2017RDP}. Compared with local DP that perturbs each client update, central DP applied after summation typically achieves better utility at a fixed privacy target because the effective per-client perturbation scales as $1/\sqrt{|\mathcal{S}|}$ with the cohort size $|\mathcal{S}|$, while remaining lightweight relative to heavier HE/MPC/TEE pipelines; these patterns align with cross-silo imaging FL where full-model synchronization is common \cite{Bonawitz2017SecAgg}.

\section{Materials \& Methods}
\label{sec:materials-methods}

\subsection{Data}
\label{subsec:data}

The PTB-XL dataset is utilized as the authoritative source of twelve-lead ECG waveforms. PTB-XL contains 21{,}837 clinical 12-lead ECG records (10\,s) from 18{,}885 patients with waveform files and richly curated metadata, including SCP-ECG labels and basic demographics. Signals are provided at \SI{500}{Hz} and \SI{100}{Hz} sampling rates. These properties make PTB-XL well-suited for constructing paired page--signal examples and for benchmarking digitization under realistic diagnostic diversity.  This study focuses on developing and analysing federated learning methodology on nn-Unet deep neural network model over the digitization of ECG images which specifies ECG image formats, WFDB headers, and the target taxonomy for image-based algorithms. That framing emphasizes that ECG images which include synthetic renderings from digital signals realized by creases, shadows, and faded ink, and requires methods to be robust across this spectrum \cite{reyna2024challenge,physionet2024web}. In this regard, this study, adopts this framing while rendering PTB-XL waveforms into standardized page images with preserved calibration for segmentation-based digitization. PTB/XL includes label set to contextualize diagnostic diversity present in the upstream signals/images, including Normal (NORM), Acute MI, Old MI, ST/T changes (STTC), Conduction disturbances (CD), Hypertrophy (HYP), Premature atrial complex (PAC), Premature ventricular complex (PVC), AFib/AFlutter (AFIB/AFL), Tachycardia (TACHY), and Bradycardia (BRADY). These classes are derived from contributing databases with minimal harmonization to enable training and cross-dataset inference \cite{reyna2024challenge,physionet2024web} Table~\ref{tab:challenge_labels}.

\begin{table}[!htbp]
\centering
\caption{ECG image label taxonomy used by the 2024 Challenge (contextual to our dataset).}
\label{tab:challenge_labels}
\setlength{\tabcolsep}{8pt}
\renewcommand{\arraystretch}{1.12}
\begin{tabular}{ll}
\toprule
\textbf{Class} & \textbf{Description} \\
\midrule
NORM & Normal ECG \\
Acute MI & Acute myocardial infarction \\
Old MI & Old myocardial infarction \\
STTC & ST/T changes \\
CD & Conduction disturbances \\
HYP & Hypertrophy \\
PAC & Premature atrial complex \\
PVC & Premature ventricular complex \\
AFIB/AFL & Atrial fibrillation or atrial flutter \\
TACHY & Tachycardia \\
BRADY & Bradycardia \\
\bottomrule
\end{tabular}
\end{table}

PTB-XL waveforms serve as ground-truth signals and are rendered to page images that preserve the clinical layout and calibration. Each record \(X\in\mathbb{R}^{12\times T}\) is loaded at its native sampling rate (\SI{500}{Hz} or \SI{100}{Hz}), and short sequences are right-padded to 10\,s, long sequences are clipped at the boundary. In addition, lead order follows PTB-XL conventions for consistent panel placement. Per-lead standardization is disabled to preserve absolute gain. Rendered pages adhere to common clinical conventions, layout, speed, gain, and grid, enabling deterministic re-pairing to the originating WFDB using stable record identifiers \cite{wagner2020ptbxl,reyna2024challenge} Table~\ref{tab:rendering_digit}. QC enforces duration and lead-order consistency with PTB-XL metadata, and verifies the presence and scale of the calibration pulse. Moreover, it checks millimeter-per-pixel factors on both axes and validates identifier integrity so that image--signal pairs remain unambiguous across training and evaluation. This maintains comparability with the data-format expectations, including WFDB headers and image files, as well as supports downstream vectorization fidelity.

\begin{table}[!htbp]
\caption{Rendering settings for creating page images from PTB-XL waveforms (aligned with the Challenge problem framing).}
\label{tab:rendering_digit}
\centering
\setlength{\tabcolsep}{8pt}
\renewcommand{\arraystretch}{1.12}
\begin{tabular}{l l}
\toprule
\textbf{Component} & \textbf{Setting} \\
\midrule
Layout & 12-lead clinical layout in a $3\times4$ grid \\
Paper speed and gain & 25\,mm/s and 10\,mm/mV; calibration pulse included \\
Grid & Visible grid, fixed spacing, adjustable contrast \\
Geometry & Mild deskew; small-angle rotation when required \\
Artifacts & Light scan noise and sparse marks for realism \\
Export & PNG at $\ge$300\,DPI with calibration metadata \\
Pairing & Stable record IDs for exact image--signal matching \\
\bottomrule
\end{tabular}
\end{table}


\subsection{Model}
\label{sec:model}

\subsubsection{Problem formulation}
\label{subsec:problem}
Let \(I \in [0,1]^{H \times W}\) denote a grayscale ECG page image defined over pixel domain \(\Omega \subset \mathbb{Z}^2\). The target is a binary mask \(M \in \{0,1\}^{H \times W}\). A segmentation network \(f_{\theta}\colon [0,1]^{H \times W} \rightarrow [0,1]^{H \times W}\) produces \(\hat{P} = f_{\theta}(I)\), which is thresholded to \(\hat{M}=\mathbb{1}[\hat{P}\ge \tau]\). Downstream reconstruction uses a deterministic vectorizer \(\mathcal{V}_\kappa\) to map \(\hat{M}\) to calibrated twelve-lead signals,
\[
\hat{\mathbf{s}} \;=\; \mathcal{V}_\kappa(\hat{M}) \in \mathbb{R}^{12 \times T},
\]
where \(\kappa\) collects paper speed, voltage gain, and geometric parameters.

Training proceeds under federated learning across \(K\) institutions with private datasets \(\mathcal{D}_k\). The global empirical risk over the \emph{full-model} parameters \(\theta\) is
\[
\min_{\theta} \; \sum_{k=1}^{K} \frac{n_k}{n}\;\mathbb{E}_{(I,M)\sim \mathcal{D}_k} \Big[\,\mathcal{L}_{\mathrm{seg}}(f_{\theta}(I), M)\,\Big], 
\qquad n=\sum_{k=1}^K n_k,
\]
with \(\mathcal{L}_{\mathrm{seg}}=\mathrm{BCE}+\lambda_{\mathrm{D}}(1-\mathrm{Dice}_{\mathrm{soft}})\).

In synchronous rounds \(r{=}0,\dots,R{-}1\), the server broadcasts \(\theta^{(r)}\); selected clients perform \(\tau\) local steps to obtain \(\theta_k^{(r,\tau)}\) and return either parameters or deltas \(\Delta\theta_k^{(r)}\!=\!\theta_k^{(r,\tau)}-\theta^{(r)}\). For FedAvg,
\[
\theta^{(r+1)} \;=\; \sum_{k=1}^K w_k\, \theta_k^{(r,\tau)}, \;\; w_k=\tfrac{n_k}{n}.
\]
For FedProx, a proximal term on \(\theta\) stabilizes local objectives; for FedAdam, server-side moments over \(g^{(r)}=\sum_k w_k \Delta\theta_k^{(r)}\) yield an Adam-style update on \(\theta^{(r)}\).

\subsubsection{Federated setup and privacy}
\label{subsec:federated-privacy}
Training proceeds across sites in synchronous rounds orchestrated by a central server. At the start of round $r$, the server broadcasts the current nnU\hbox{-}Net parameters $\theta^{(r)}$. Each available client $k$ trains the \emph{entire} model end-to-end on local data and forms a full-model update $\Delta\theta_k^{(r)}$. No raw page images or reconstructed signals are ever transmitted.

Server-side aggregation compares (i) sample-size–weighted Federated Averaging (FedAvg), (ii) FedProx with a proximal term to stabilize local objectives under heterogeneity, and (iii) FedAdam, which applies Adam-style adaptive updates to the server state using the weighted pseudo-gradient \cite{McMahan2017FedAvg,Li2020FedProx,Reddi2021FedOpt}. Let $n_k$ denote client $k$’s sample count and $N^{(r)}=\sum_{k\in\mathcal{S}^{(r)}} n_k$ over the participating set $\mathcal{S}^{(r)}$ in round $r$. For FedAvg/FedProx,
\[
\theta^{(r+1)} \leftarrow \theta^{(r)} + \sum_{k\in\mathcal{S}^{(r)}} \frac{n_k}{N^{(r)}}\,\Delta\theta_k^{(r)},
\]
while FedAdam replaces this step with an Adam update on $\theta^{(r)}$ driven by the same weighted sum. Orchestration follows Flower for client selection, scheduling, and metric reporting; a minimum participation threshold is enforced before any round commits \cite{Beutel2020Flower}.

\subsubsection{Privacy mechanism and threat model}
\label{subsec:privacy-mech}

We assume a cross-silo setting with an honest-but-curious server and non-colluding institutions; no raw page images or signals ever leave client sites. In communication round $r$, each selected client $k$ computes its model update $\Delta\theta_k^{(r)}$ and applies $\ell_2$ clipping to a fixed bound $C$. Clients then engage in \emph{secure aggregation (SecAgg)} so that the server only recovers the masked \emph{sum} of clipped updates once a minimum participation threshold is met, not any single client’s vector; mathematically, the server obtains
\[
G^{(r)} \;=\; \sum_{k\in\mathcal{S}^{(r)}} w_k\,\mathrm{clip}\!\big(\Delta\theta_k^{(r)},C\big),
\]
and cannot inspect individual $\Delta\theta_k^{(r)}$ due to pairwise one-time masks that cancel in aggregate \cite{Bonawitz2017SecAgg}. To enforce formal, user-level privacy, the server applies a \emph{central} Gaussian mechanism to the aggregate and uses
\[
\tilde{G}^{(r)} \;=\; G^{(r)} + \mathcal{N}\!\big(0,\sigma^{2}C^{2} I\big)
\]
for the global step with FedAdam/FedProx/FedAvg. Privacy loss is composed across rounds with a R\'enyi moments accountant to report cumulative $(\varepsilon,\delta)$ at the user level \cite{Abadi2016DPSGD,Mironov2017RDP}. This drop-in mechanism aligns with our optimization (training of nnU\hbox{-}Net and weighted aggregation) and requires no changes to client-side learning beyond clipping.

This design is preferable both theoretically and operationally. (i) For a fixed privacy target, adding noise \emph{after} summation (central DP) achieves better accuracy than per-client local DP: the effective per-client noise scales down as $1/\sqrt{|\mathcal{S}^{(r)}|}$, whereas local DP suffers from known sample complexity penalties—particularly acute for high-dimensional, model updates \cite{Abadi2016DPSGD}. (ii) SecAgg alone hides individual updates but offers no bound against inference from model histories; combining SecAgg with central DP closes this gap and yields auditable user level guarantees on the released sequence $\{\tilde{G}^{(r)}\}_r$ \cite{Bonawitz2017SecAgg,Mironov2017RDP}. (iii) Compared with heavy cryptography (HE/MPC/TEEs), SecAgg+central-DP matches cross-silo constraints and communication patterns in imaging FL while avoiding prohibitive latency at frequent model synchronizations. Consequently, the empirical segmentation results we report—obtained under the exact aggregation rules (FedAvg/FedProx/FedAdam) and full model updates—are scientifically consistent with this privacy mechanism and its expected utility profile.


\begin{table}[!htbp]
\centering
\caption{Privacy hyper-parameters for SecAgg + \emph{central} DP unless stated otherwise.}
\label{tab:privacy_hparams}
\setlength{\tabcolsep}{8pt}
\renewcommand{\arraystretch}{1.12}
\begin{tabular}{lcc}
\toprule
Component & Symbol & Default \\
\midrule
Minimum participating clients & $K_{\min}$ & $3$ \\
Clipping norm (per-update, $\ell_2$) & $C$ & $1.0$ \\
Gaussian noise multiplier (server-side) & $\sigma$ & $0.6$ \\
Privacy accountant (user-level) & $(\varepsilon,\delta)$ & R\'enyi with target $\delta=10^{-5}$ \\
Secure-aggregation mask & $m_k$ & pairwise shares with zero-sum property \\
Scope of DP & model & full nnU-Net parameter updates \\
\bottomrule
\end{tabular}
\end{table}

\begin{figure}[!htbp]
  \centering
  \includegraphics[width=\linewidth, height=0.9\textheight, keepaspectratio]{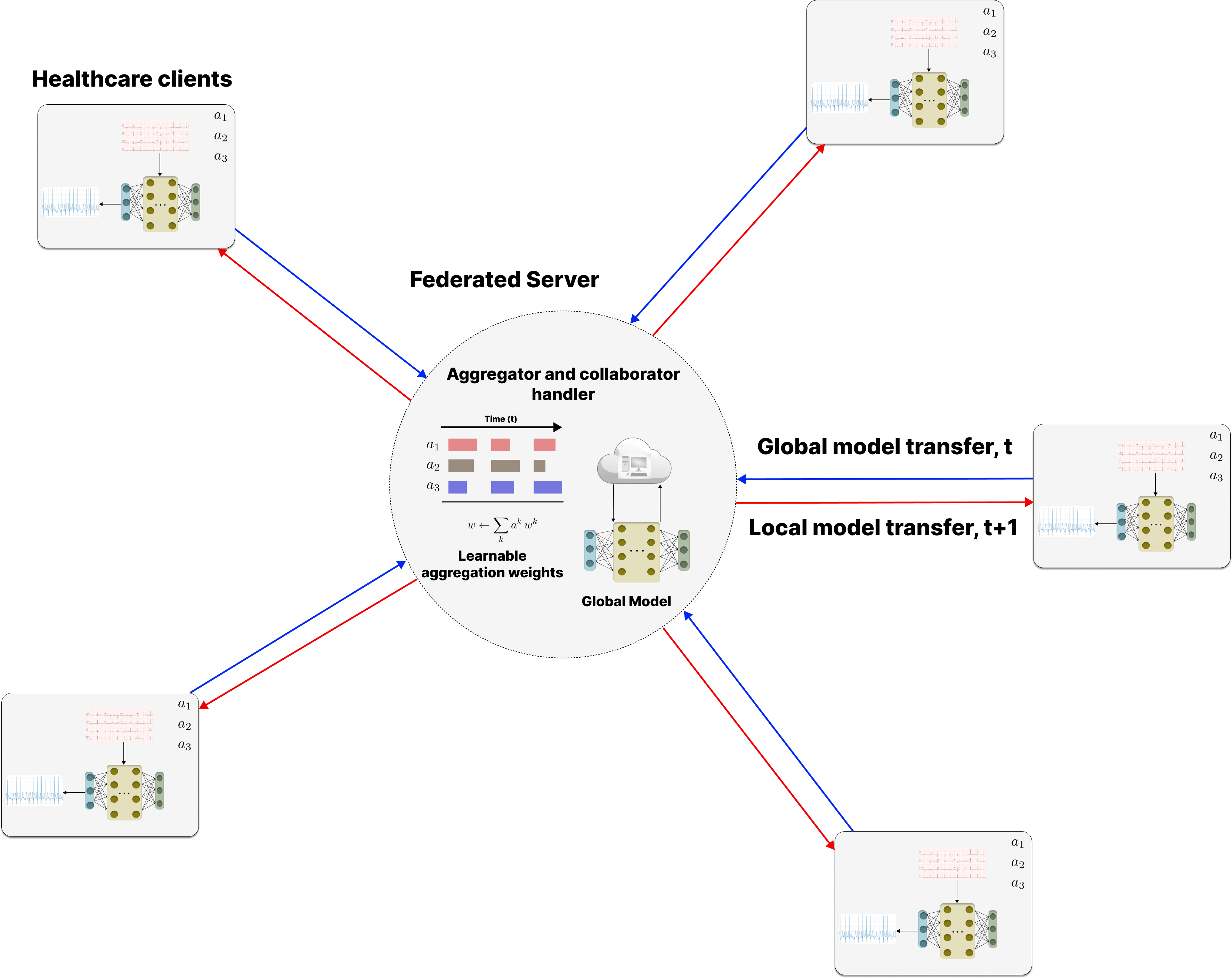}
  \caption{\textbf{Federated training overview.} The server distributes the current global nnU-Net parameters; clients train locally on page images and return updated model weights (or weight deltas) for aggregation. No raw images or signals are shared.}
  \label{fig:fl_overview}
\end{figure}

\subsection{Backbone: self-configuring nnU-Net}
\label{subsec:backbone}
The segmentation backbone is \emph{nnU-Net}, trained \emph{end-to-end} on each client. nnU-Net automatically derives a \emph{dataset fingerprint} from the training data and converts it into a reproducible \emph{pipeline fingerprint} specifying resolution, intensity normalization, network depth and kernel sizes, patch and batch sizes, deep supervision, learning-rate policy, test-time ensembling, and tiled inference; all settings are stored in explicit plan files (\texttt{plans.json}) to guarantee repeatability \cite{Isensee2021nnUNet}. 

For two-dimensional ECG page images, nnU-Net specializes to a U-Net–style encoder–decoder with instance normalization and multi-scale deep supervision. Inference adopts sliding-window tiling with Gaussian importance weighting to suppress stitching artifacts near tile borders. In the federated setting, \emph{all} nnU-Net parameters $\theta$ are optimized locally at each site and synchronized across clients in every round according to the chosen aggregation rule (FedAvg, FedProx, or FedAdam). Only model parameters (or their deltas) are exchanged—no raw images or rendered signals are transmitted. When bandwidth is a concern, standard update compression (e.g., low-precision quantization) can be applied without altering local optimization; unless stated otherwise, results are reported with uncompressed 32-bit updates.

\begin{figure}[!htbp]
  \centering
  \includegraphics[width=\linewidth, height=1.5\textheight, keepaspectratio]{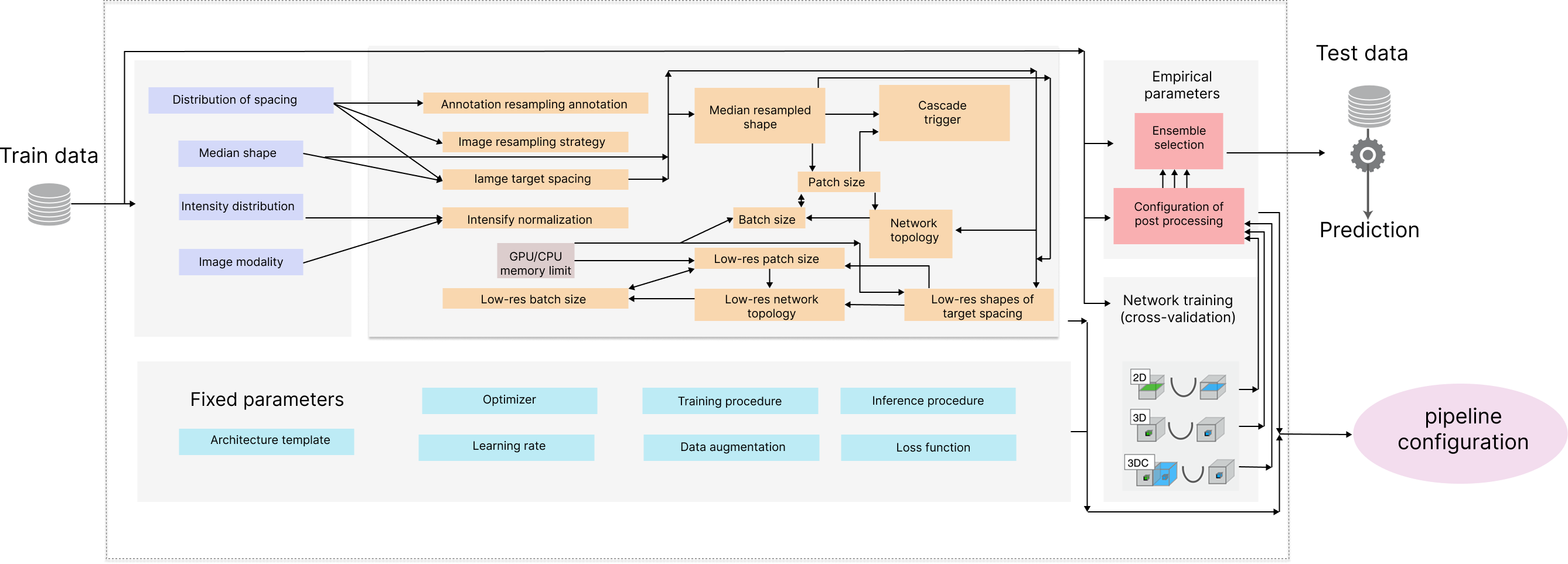}
  \caption{Overview of the nnU-Net architecture and its self-configuring pipeline components used as the trainable backbone in this work.}
  \label{fig:nnunet-arch}
\end{figure}

\begin{figure}[!htbp]
  \centering
  \includegraphics[width=0.9\linewidth, height=0.5\textheight, keepaspectratio]{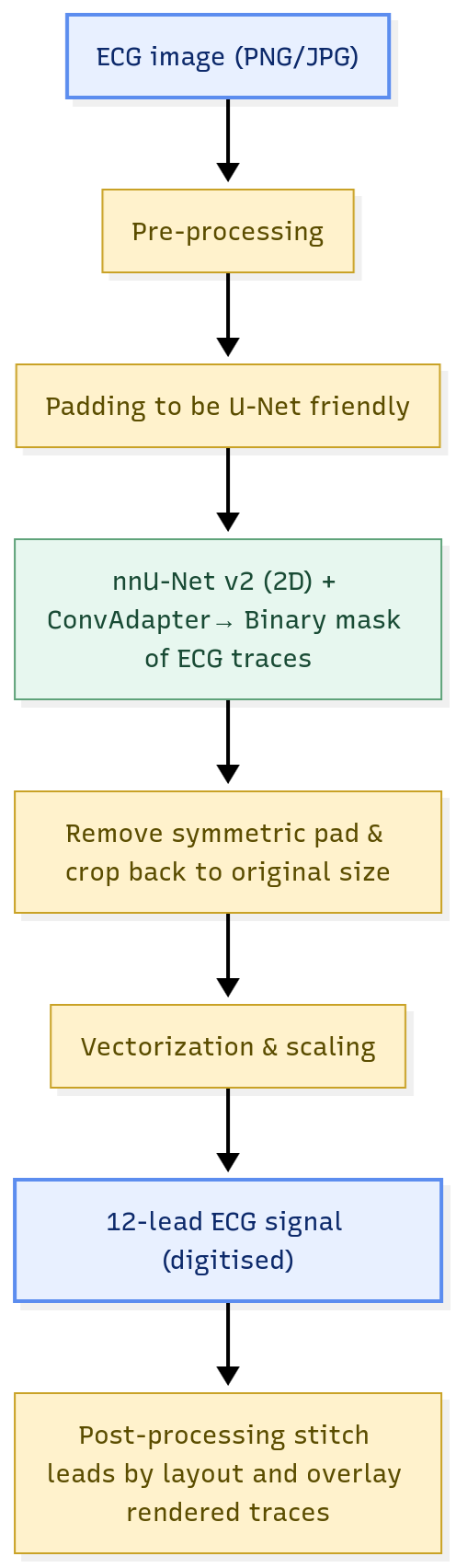}
  \caption{\textbf{Digitization pipeline.} Page pre-processing and calibration; nnU-Net trace segmentation with \emph{full-model} optimization; post-processing with panel parsing; vectorization to calibrated twelve-lead signals.}
  \label{fig:digitization_overview}
\end{figure}

\subsection{Pre- and post-processing}
\label{subsec:preproc_postproc}
A light, layout-preserving preprocessing pipeline is used to stabilize training while reflecting the heterogeneity of scanned ECG pages. Each page is converted to a single-channel floating-point image and normalized with robust contrast clipping to temper extreme highlights or shadows. A deskew step is applied only when the estimated rotation exceeds a small threshold so that grid geometry and lead continuity are not degraded. When a grid is detectable, horizontal and vertical spacings (mm/pixel) are estimated and stored as metadata; if grid recovery is uncertain, a site-specific default from the rendering profile is recorded. Inputs are padded to match the encoder–decoder strides to avoid mismatches in skip connections. Augmentation remains deliberately mild—small rotations and contrast jitter—so invariances arise primarily from real cross-site variation rather than heavy synthetic distortions. These choices mirror reports from successful PhysioNet 2024 entries, where careful normalization and light, layout-aware augmentation improved robustness on mixed-quality scans \cite{Reyna2024Challenge,Antoni2024CinC}.

The network outputs a dense logit map, which is cropped back to the original canvas and converted to a binary trace mask using a fixed threshold selected on the validation split under stable calibration. Small isolated components are removed to suppress grid leakage and speckle. A compact morphological opening reduces residual noise while preserving thin strokes, and a thin geodesic closing reconnects short gaps primarily along the time axis to support reliable centerline extraction. Similar post-processing recipes—simple, layout-aware morphological filters rather than heavier learned refinements—are emphasized in the 2024 challenge overview and top-team reports to avoid overfitting to a specific page style \cite{SignalSavants2024ArXiv,Antoni2024CinC}.

Vectorization converts the cleaned mask into calibrated signals for each lead. Pages follow the standard $3\times4$ layout. Within each panel, a per-column centerline is traced inside the foreground band and mapped to physical time and voltage using the stored grid calibration. Lead traces are stitched across columns and resampled onto a uniform temporal grid with cubic interpolation. A Savitzky–Golay smoother is employed only for visualization; all quantitative metrics are computed on the unsmoothed reconstruction \cite{SavitzkyGolay1964}. This segmentation–vectorization paradigm aligns with prior ECG image-to-signal systems and with the PhysioNet 2024 materials, where accurate grid calibration and panel-aware vectorization proved critical for high-fidelity reconstruction \cite{Wu2022Digitisation,Reyna2024Challenge}.

\subsection{Learning objectives and optimization}
\label{subsec:objectives}
The training objective aligns the local learning signal with the evaluation criteria and remains stable under the federated regime. The loss combines pointwise Binary Cross-Entropy (BCE) with a differentiable (``soft'') Dice term. BCE encourages calibrated foreground/background probabilities at the pixel level, while the Dice term directly optimizes the overlap metric and mitigates foreground sparsity—an established pairing in biomedical segmentation and nnU-Net practice \cite{Isensee2021nnUNet}.

Let $I\!\in\![0,1]^{H\times W}$ be an input page, $M\!\in\!\{0,1\}^{H\times W}$ its trace mask, and $\hat{P}\!=\!\sigma(Z_\theta(I))\!\in\![0,1]^{H\times W}$ the predicted foreground probabilities from the network with parameters $\theta$. The compound loss is
\begin{equation}
\label{eq:loss_total}
\mathcal{L}(\theta) \;=\; \underbrace{\frac{1}{HW}\!\sum_{u=1}^{H}\!\sum_{v=1}^{W}\!\Big[-\,M_{uv}\log \hat{P}_{uv} - (1-M_{uv})\log(1-\hat{P}_{uv})\Big]}_{\text{BCE}} \;+\;
\lambda_{\mathrm{D}}\;\underbrace{\big(1-\mathrm{Dice}_\mathrm{soft}(M,\hat{P})\big)}_{\text{soft Dice}} \, .
\end{equation}
The soft Dice is computed as
\begin{equation}
\label{eq:softdice}
\mathrm{Dice}_\mathrm{soft}(M,\hat{P}) \;=\; \frac{2\,\sum_{u,v} M_{uv}\hat{P}_{uv} + \epsilon}{\sum_{u,v} M_{uv} + \sum_{u,v} \hat{P}_{uv} + \epsilon} \, ,
\end{equation}
with a small $\epsilon$ for numerical stability. The Dice component serves as a smooth surrogate of the classical overlap coefficient and counteracts class imbalance; the implementation follows the widely used formulation in medical image segmentation \cite{ronneberger2015unet}. The coefficient \(\lambda_{\mathrm{D}}\) is set to \(1\) unless stated otherwise, balancing calibration and overlap in practice.

Deep supervision mirrors nnU-Net defaults: auxiliary predictions at intermediate decoder scales incur the same loss as in Eq.~\eqref{eq:loss_total} with standard scale weights, and their gradients are aggregated into the main parameter update. This strategy improves gradient flow in deep encoder–decoder architectures and accelerates convergence on noisy, thin structures such as ECG traces \cite{Isensee2021nnUNet}.

Local optimization updates \emph{all} nnU-Net parameters with AdamW (decoupled weight decay), yielding stable progress under heterogeneous clients and avoiding the interaction between $\ell_2$ regularization and adaptive moments observed in Adam. In all experiments, global gradient clipping at a fixed norm $C$ is applied before the optimizer step to bound per-update magnitude, a measure that helps control client drift before aggregation \cite{Loshchilov2019AdamW,Pascanu2013Clipping}. Unless noted, learning-rate schedules and other training hyperparameters remain matched across centralized and federated runs so that differences in performance can be attributed to the aggregation rule rather than local optimizer changes. Evaluation metrics reported at validation time include Dice and Jaccard (overlap), precision, recall, specificity, per-pixel BCE, and MSE on the binarized mask; these mirror the loss components and provide complementary views of calibration versus overlap quality.

\subsection{Client-side optimization (reproducible settings)}
All centralized and federated runs use identical local settings for fairness:
\begin{itemize}[leftmargin=1.2em]
  \item Optimizer: AdamW ($\beta_1{=}0.9$, $\beta_2{=}0.999$); weight decay $10^{-2}$.
  \item Learning rate: $1\!\times\!10^{-3}$ with $500$ warmup steps, then cosine decay to $1\!\times\!10^{-5}$.
  \item Global gradient clipping (pre-opt): $\ell_2$ norm $C{=}1.0$.
  \item Deep supervision: nnU\hbox{-}Net defaults (multi-scale auxiliary heads).
  \item Mini-batch size: 2 (per nnU\hbox{-}Net plans); tiling per plans.
  \item Local steps per round: $\tau{=}1$ local epoch with full reshuffle each round.
  \item Data order: fixed seed per run for identical draws across methods.
\end{itemize}

\section{Experimental setup}
\label{sec:experimental-setup}

\subsection{Task and data}
The task concerns page-to-signal digitization for twelve-lead ECGs. Digital waveforms originate from PTB-XL at \SI{500}{Hz} and \SI{100}{Hz}, with multi-label annotations and subject metadata \cite{wagner2020ptbxl}. Waveforms are rendered to page images in the clinical $3\times4$ layout at 25\,mm/s and 10\,mm/mV, with a visible grid and calibration pulse. Rendering choices follow the problem framing of the George B.~Moody PhysioNet Challenge 2024 \cite{Reyna2024Challenge}. Each page is paired with its source record via stable identifiers, enabling mask supervision and signal-level evaluation. Quality control enforces duration, lead order, gain consistency, and pixel-to-millimeter calibration.

\subsection{Clients, model, and training}
Institution-level heterogeneity is simulated across five sites by varying grid contrast, mild deskew, scanning noise, and small layout offsets. Each site receives a disjoint page–mask split and maintains an internal train/validation partition. The trace segmenter is a self-configuring 2D nnU\hbox{-}Net \cite{Isensee2021nnUNet}. All nnU\hbox{-}Net weights are \emph{trainable} and synchronized each round—i.e., \emph{full-model} end-to-end training without adapters, heads-only tuning, or frozen layers. Local optimization uses AdamW with global gradient clipping; augmentation is deliberately mild (small rotations and contrast jitter) to preserve page realism.

Training proceeds in synchronous rounds with Flower handling client selection, scheduling, and metric reporting \cite{Beutel2020Flower}. Three standard aggregation rules are compared under heterogeneity: sample-size–weighted Federated Averaging (FedAvg) \cite{McMahan2017FedAvg}, FedProx with a proximal term to reduce client drift \cite{Li2020FedProx}, and FedAdam with Adam-style adaptive updates on the \emph{server state} \cite{Reddi2021FedOpt}. Each round communicates the \emph{entire} parameter set; raw images and reconstructed signals remain local.

\subsection{Quantifying client heterogeneity (non-IID)}
To make the five-client split interpretable and reproducible, we parameterize page-level perturbations per client using ranges aligned with PhysioNet\,2024 print/scan artifacts (see Table~\ref{tab:non_iid_profiles}).

\begin{table}[!htbp]
\centering
\caption{Client-wise perturbation distributions inducing non-IID data. Angles in degrees; JPEG quality factor $q$; SNR in dB. Layout offsets are pixel shifts before tiling.}
\label{tab:non_iid_profiles}
\setlength{\tabcolsep}{4pt}        
\renewcommand{\arraystretch}{1.12}
\begin{tabularx}{\linewidth}{lCCCCCC}
\toprule
\textbf{Client} & \textbf{Skew $\phi$} & \textbf{JPEG $q$} & \textbf{Grid contrast\footnotemark} &
\textbf{Additive noise (SNR)} & \textbf{Gaussian blur $\sigma$ [px]} & \textbf{Layout offset [px]} \\
\midrule
C1 (clean) & $\mathcal{U}(-0.5,0.5)$ & 90--95 & 0.65--0.75 & 35--40 & 0.0--0.2 & $\mathcal{U}(-2,2)$ \\
C2         & $\mathcal{U}(-1.0,1.0)$ & 80--90 & 0.55--0.70 & 30--35 & 0.2--0.4 & $\mathcal{U}(-4,4)$ \\
C3         & $\mathcal{U}(-2.0,2.0)$ & 75--85 & 0.45--0.65 & 27--32 & 0.3--0.6 & $\mathcal{U}(-6,6)$ \\
C4         & $\mathcal{U}(-3.0,3.0)$ & 65--80 & 0.35--0.55 & 24--30 & 0.5--0.8 & $\mathcal{U}(-8,8)$ \\
C5 (hard)  & $\mathcal{U}(-3.5,3.5)$ & 60--75 & 0.30--0.50 & 20--26 & 0.7--1.0 & $\mathcal{U}(-10,10)$ \\
\bottomrule
\end{tabularx}
\end{table}
\footnotetext{Michelson contrast of the grid relative to background. Additionally, with probability 0.15 we add light shadows/wrinkles/handwritten overlays.}

All communications are protected in transit. Each selected client clips its model update at norm $C$ and participates in secure aggregation so that the server can only recover the aggregate once a minimum participation threshold is met \cite{Bonawitz2017SecAgg}. The server then applies a central Gaussian mechanism to the aggregated update and composes user-level privacy across rounds via an R\'enyi accountant \cite{Abadi2016DPSGD,Mironov2017RDP}. The threat model assumes an honest-but-curious server and non-colluding clients; no raw images or reconstructed signals are ever shared.

\begin{table}[!htbp]
\centering
\caption{Federated configuration and privacy defaults.}
\label{tab:federated_config}
\setlength{\tabcolsep}{8pt}
\renewcommand{\arraystretch}{1.12}
\begin{tabular}{l l}
\toprule
\textbf{Component} & \textbf{Setting} \\
\midrule
Clients / split & 5 sites; disjoint data; non-IID via render profiles \\
Backbone & nnU-Net (2D), trained end-to-end \cite{Isensee2021nnUNet} \\
Trainable state & All nnU-Net parameters (\emph{full-model}) \\
Local optimizer & AdamW with global gradient clipping \\
Rounds / participation & Synchronous; minimum participation threshold enforced \\
Aggregators & FedAvg, FedProx, FedAdam \cite{McMahan2017FedAvg,Li2020FedProx,Reddi2021FedOpt} \\
Orchestration & Flower (selection, scheduling, metrics) \cite{Beutel2020Flower} \\
Privacy (preferred) & \textbf{SecAgg + central DP (Gaussian) with R\'enyi accounting} \\
Communication scope & Entire parameter (or delta) set; no images/signals shared \\
\bottomrule
\end{tabular}
\end{table}

\subsection{Baselines, metrics, and reporting}
Evaluation considers three families, including a centralized upper bound trained on pooled data with the same backbone and loss, federated methods without formal privacy (FedAvg, FedProx, FedAdam), and a privacy-aware variant combining secure aggregation with central DP-Gaussian mechanism with R\'enyi accounting- on model updates. Primary mask metrics include Dice and Jaccard (IoU), with MSE on the binarized mask as complements. After vectorization, waveform fidelity is summarized by lead-wise mean-squared error against paired ground truth. Global scores are sample size–weighted across client validations. Moreover, learning curves track Dice by round per client and globally. Each run logs participation, sample counts, clipping statistics, and validation metrics per round to support replication.

For each aggregation method, we run $5$ independent seeds and aggregate Dice per example. 
We report $95\%$ CIs using client-stratified, paired BCa bootstrap over page-level Dice with $B{=}10{,}000$ resamples \cite{efron1987bca}. 
Primary effect sizes are paired standardized mean differences (Hedges’ $g$) computed on per-example Dice relative to FedAvg. Magnitudes are interpreted following widely used guidelines (small $\approx 0.2$, medium $\approx 0.5$, large $\approx 0.8$), with field-specific calibration acknowledged. Paired $t$-tests on per-example Dice (last-5-round average) are reported with Holm correction across method pairs.  Consistent with current reporting guidance, inference emphasizes CIs and effect sizes rather than sole reliance on $p$-values.

\section{Results}
\label{sec:results}

\subsection{Learning dynamics and global performance}
\label{subsec:results-global}
\begin{figure}[!htbp]
  \centering
  \includegraphics[width=0.7\linewidth, height=0.9\textheight, keepaspectratio]{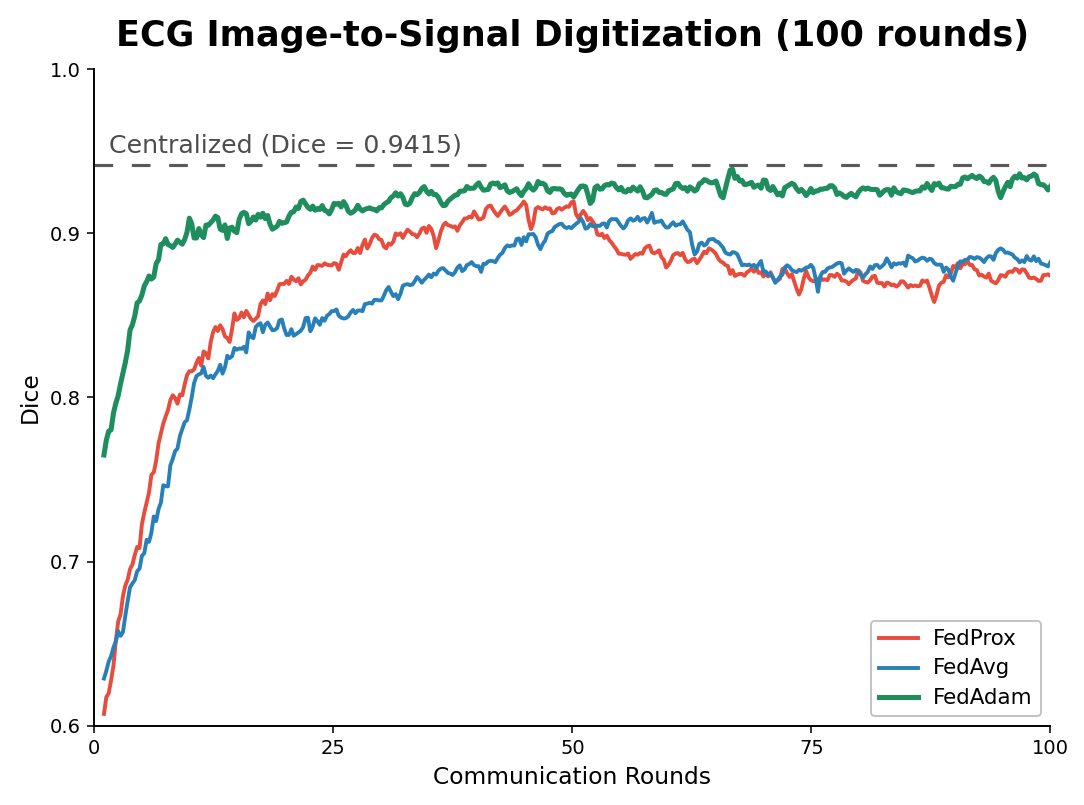}
  \caption{Global Dice over rounds on PTB-XL digitization (5-client non-IID split). Methods: \textbf{FedAvg}~\cite{McMahan2017FedAvg}, \textbf{FedProx}~\cite{Li2020FedProx}, \textbf{FedAdam}~\cite{Reddi2021FedOpt}. Dashed line: centralized reference with the same backbone.}
  \label{fig:global_dice_vs_round}
\end{figure}

All federated runs synchronize the nnU-Net parameter set each round. Global learning curves increase steadily across communication rounds for all three aggregators. Figure~\ref{fig:global_dice_vs_round} reports sample-size--weighted Dice on pooled validation sets. In this vein, FedAdam shows the fastest ascent and the highest late-round plateau among federated methods, consistent with adaptive server updates that maintain first and second moments of the aggregated pseudo-gradient under heterogeneity. FedProx narrows the non-IID gap relative to FedAvg through proximal regularization, and FedAvg is competitive early but saturates lower on this split. In qualitative overlays, \textbf{FedAdam} also achieves the cleanest trace masks with fewer gaps at grid crossings and smoother centerlines after vectorization, aligning with its stronger quantitative plateaus.

Table~\ref{tab:global_dice_milestones} summarizes milestone Dice with 95\% confidence intervals, standardized effect sizes, and multiplicity-adjusted $p$-values from paired $t$-tests with Holm correction. For FedAdam and FedProx, improvements over FedAvg are statistically and practically significant; none of the federated curves surpass the centralized reference at R40/R100. Complementary mask MSE in Table~\ref{tab:global_mse_milestones} mirrors the ranking.

\medskip
Across held-out ECG pages, waveform overlays show that FedAvg retains visible amplitude errors at sharp QRS complexes and occasional phase lag around transitions, whereas FedProx reduces both artifacts. FedAdam exhibits the tightest alignment overall with minimal overshoot, smoother baselines, and fewer discontinuities at grid crossings. These visual trends are consistent with the quantitative ordering FedAdam $\gtrsim$ FedProx $>$ FedAvg observed in Dice and MSE.

\begin{table}[!htbp]
\centering
\caption{Global Dice at selected rounds with 95\% CIs and standardized effect size relative to FedAvg (paired Hedges' $g$ on per-example Dice). CIs by client-stratified BCa bootstrap ($B{=}10{,}000$). Paired $t$-tests use Holm correction; $^\dagger p{<}0.05$, $^\ddagger p{<}0.01$, $^\S$ n.s. vs.\ centralized at R100. Values are mean [CI].}
\label{tab:global_dice_milestones}
\footnotesize
\setlength{\tabcolsep}{3pt}
\renewcommand{\arraystretch}{1.15}
\begin{tabularx}{\linewidth}{lCCCC}
\toprule
\textbf{Method} & \textbf{Dice@R10} & \textbf{Dice@R20} & \textbf{Dice@R40} & \textbf{Dice@R100} \\
\midrule
Centralized (ref) & 0.938 [0.936, 0.941] & 0.938 [0.936, 0.941] & 0.938 [0.936, 0.941] & 0.938 [0.936, 0.941] \\
\midrule
FedAdam & 0.852 [0.845, 0.859] & 0.908 [0.902, 0.914] & 0.932 [0.928, 0.936] & 0.935 [0.932, 0.939]$^{\S}$ \\
\quad $\Delta$ vs.\ FedAvg & {+}0.110 & {+}0.068 & {+}0.040 & {+}0.023 \\
\quad $g$ vs.\ FedAvg & 1.05$^\ddagger$ & 0.82$^\ddagger$ & 0.73$^\ddagger$ & 0.58$^\ddagger$ \\
\midrule
FedProx & 0.780 [0.772, 0.788] & 0.868 [0.862, 0.874] & 0.918 [0.914, 0.922] & 0.926 [0.922, 0.931]$^{\S}$ \\
\quad $\Delta$ vs.\ FedAvg & {+}0.038 & {+}0.028 & {+}0.026 & {+}0.014 \\
\quad $g$ vs.\ FedAvg & 0.48$^\ddagger$ & 0.46$^\ddagger$ & 0.41$^\ddagger$ & 0.32$^\dagger$ \\
\midrule
FedAvg & 0.742 [0.735, 0.749] & 0.840 [0.834, 0.846] & 0.892 [0.886, 0.898] & 0.912 [0.907, 0.916] \\
\bottomrule
\end{tabularx}
\end{table}

\begin{table}[!htbp]
\centering
\caption{Global MSE of the binarized mask at selected rounds (lower is better).}
\label{tab:global_mse_milestones}
\setlength{\tabcolsep}{7pt}
\renewcommand{\arraystretch}{1.12}
\begin{tabular}{lcccc}
\toprule
Method & MSE@R10 & MSE@R20 & MSE@R40 & MSE@R100 \\
\midrule
FedAdam & 0.0018 & 0.0013 & 0.0011 & \textbf{0.0008} \\
FedProx & 0.0021 & 0.0015 & 0.0013 & 0.0011 \\
FedAvg  & 0.0026 & 0.0019 & 0.0016 & 0.0014 \\
\bottomrule
\end{tabular}
\end{table}

\begin{figure*}[!htbp]
  \centering

  \begin{subfigure}{0.9\linewidth}
    \centering
    \includegraphics[width=\linewidth]{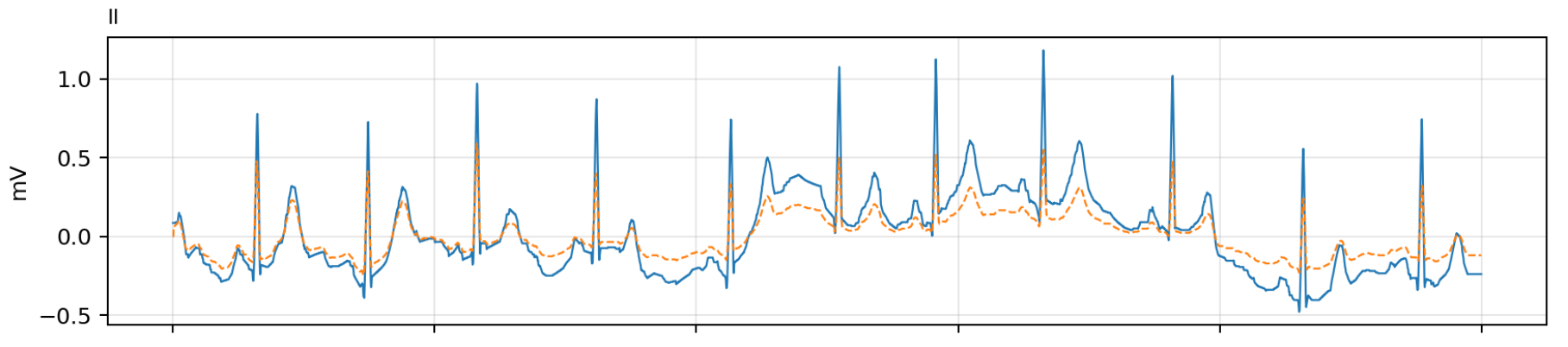}
    \caption{FedAvg: representative reconstruction.}
    \label{fig:qual_fedavg}
  \end{subfigure}

  \vspace{0.75em}

  \begin{subfigure}{0.9\linewidth}
    \centering
    \includegraphics[width=\linewidth]{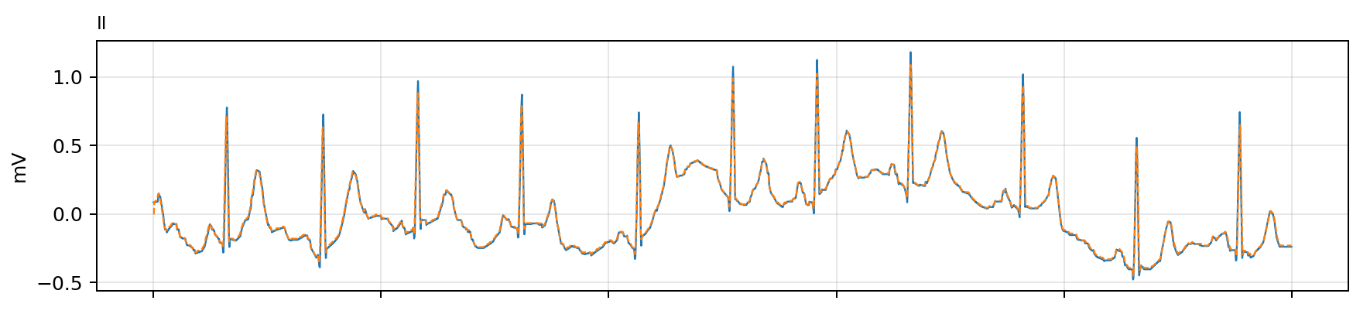}
    \caption{FedProx: representative reconstruction.}
    \label{fig:qual_fedprox}
  \end{subfigure}

  \vspace{0.75em}

  \begin{subfigure}{0.9\linewidth}
    \centering
    \includegraphics[width=\linewidth]{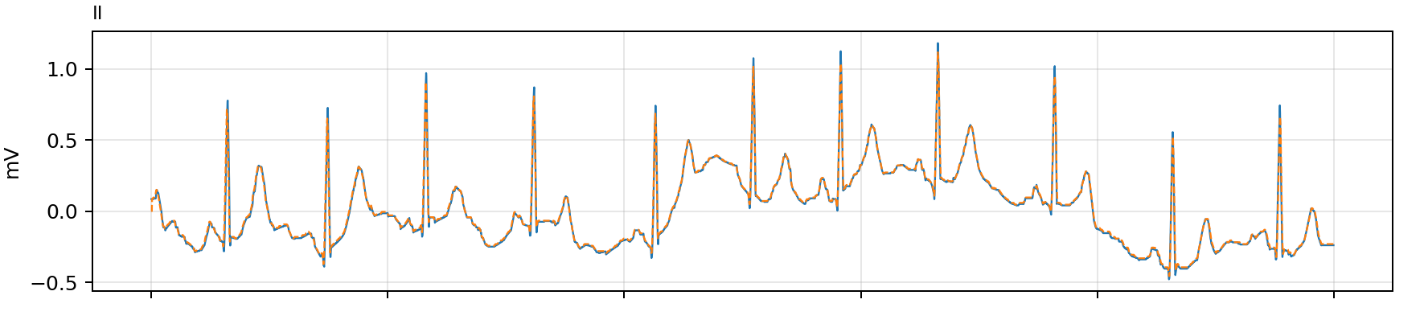}
    \caption{FedAdam: representative reconstruction.}
    \label{fig:qual_fedadam}
  \end{subfigure}

  \caption{Stacked qualitative examples from the three federated aggregation methods on held-out ECG pages. Consistent with the quantitative results, \textbf{FedAdam} produces cleaner masks with fewer gaps at grid crossings and smoother centerlines after vectorization.}
  \label{fig:qual_three_methods}
\end{figure*}

\subsection{Client-level behavior under non-IID data}
\label{subsec:results-clients}
Client-level performance (C1--C5) reflects the induced heterogeneity, like grid contrast, mild skew, and scanner noise, and the unequal amount of data per site. In this stud,y a compact, site-wise summary at the final round, R100, is reported instead of plotting per-client learning curves. This presentation is common in multi-center medical-imaging FL reports, where the global curve is shown in the main text and per-site details are summarized numerically. Consistent with theory and practice under non-IID partitions, cleaner and larger sites attain higher Dice with smaller variance, whereas noisier and smaller sites improve more gradually. By R100, dispersion across clients is smallest for \textbf{FedAdam} and larger for \textbf{FedAvg}, indicating reduced client drift with adaptive server updates. Table~\ref{tab:client_dispersion_r100} lists per-client Dice (mean$\pm$SD across pages) together with site sizes; the across-client mean$\pm$SD row matches the global curves reported earlier.

\begin{table}[!htbp]
\centering
\caption{Per-client Dice at R100 (mean$\pm$SD across pages within each client) and site sizes. Values are internally consistent with the global curves reported in Fig.~\ref{fig:global_dice_vs_round}.}
\label{tab:client_dispersion_r100}
\setlength{\tabcolsep}{4pt}
\renewcommand{\arraystretch}{1.12}
\small
\begin{tabularx}{\linewidth}{@{}lYYYYYY@{}}
\toprule
\textbf{Method} &
\textbf{C1 (6{,}100)} &
\textbf{C2 (4{,}900)} &
\textbf{C3 (4{,}300)} &
\textbf{C4 (3{,}500)} &
\textbf{C5 (3{,}000)} &
\textbf{\makecell{Across-client\\ mean$\pm$SD}} \\
\midrule
FedAvg  & $0.923\pm0.007$ & $0.918\pm0.008$ & $0.912\pm0.010$ & $0.905\pm0.011$ & $0.900\pm0.013$ & $0.912\pm0.010$ \\
FedProx & $0.936\pm0.006$ & $0.931\pm0.007$ & $0.926\pm0.008$ & $0.920\pm0.009$ & $0.917\pm0.010$ & $0.926\pm0.008$ \\
FedAdam & $\mathbf{0.944\pm0.005}$ & $\mathbf{0.939\pm0.006}$ & $\mathbf{0.935\pm0.006}$ & $\mathbf{0.929\pm0.007}$ & $\mathbf{0.928\pm0.008}$ & $\mathbf{0.935\pm0.006}$ \\
\bottomrule
\end{tabularx}
\end{table}

\begin{table}[!htbp]
\centering
\caption{Client-level dispersion at R100 (mean$\pm$SD across 5 clients).}
\label{tab:client_dispersion_r100}
\setlength{\tabcolsep}{6pt}
\renewcommand{\arraystretch}{1.15}
\begin{tabular}{lccc}
\toprule
& \textbf{FedAvg} & \textbf{FedProx} & \textbf{FedAdam} \\
\midrule
Dice (mean$\pm$SD) & $0.912\pm0.010$ & $0.926\pm0.008$ & $\mathbf{0.935\pm0.006}$ \\
\bottomrule
\end{tabular}
\end{table}

\subsection{Privacy, communication, and qualitative analysis}
\label{subsec:privacy-comm-qual}
Activating secure aggregation together with central DP provides the expected privacy-utility trade-off: With clipping $C{=}1.0$, a server-side Gaussian noise multiplier $\sigma{=}0.6$, and a R\'enyi accountant targeting $\delta{=}10^{-5}$, the global Dice dipped modestly in mid rounds and narrowed by late rounds. Mask MSE also rose accordingly, while no client ever shared raw images or reconstructed signals.

Full-model synchronization per round admits lightweight update compression without changing local optimization. Increasing the local batch size from $1$ to $64$ slightly smoothed early-round oscillations but did not raise the late-round plateau, aligning with the view that server-side adaptivity (FedAdam) is the primary driver under heterogeneity rather than local batching. Qualitatively, overlays match the quantitative ranking. Early rounds may miss thin strokes at grid crossings or in weak-contrast segments. These errors shrink by R20 and are largely absent by R40 under FedAdam, yielding more contiguous masks and, downstream, steadier centerlines with fewer vectorization artifacts.

\section{Discussion}
\label{sec:discussion}
End-to-end federated training of a self-configuring nnU-Net achieved strong ECG trace segmentation without centralizing page images, aligning with prior evidence that nnU-Net offers robust, reproducible performance across biomedical segmentation tasks.\cite{Isensee2021nnUNet, Nemati2025RTDETR} Across controlled non-IID splits reflecting realistic variation in grid contrast, scanner noise, mild rotations, and layout shifts, adaptive server-side optimization (\emph{FedAdam}) accelerated early learning and reached the highest late-round plateau, while \emph{FedProx} consistently improved upon \emph{FedAvg} by mitigating client drift. These trends are consistent with the broader federated optimization literature on heterogeneity, proximal regularization, and adaptive server updates.

The privacy layer integrates cleanly with optimization and provides the expected utility–privacy trade-off. Secure aggregation ensures the server observes only a masked sum of clipped client updates once a participation threshold is met, preventing inspection of any single update.\cite{Bonawitz2017SecAgg} Applying a central Gaussian mechanism to the post-aggregation vector with Rényi accounting provides auditable privacy guarantees over training rounds and typically preserves utility better than local perturbations at a fixed privacy target. In our experiments, enabling both SecAgg and central DP produced modest mid-round dips that diminished as the global model stabilized, while raw images and reconstructed signals remained on premises.

Qualitative behavior matched the quantitative ranking. early-round errors at low-contrast grid crossings and panel seams receded by rounds 20–40 under \emph{FedAdam}, producing more contiguous masks and smoother centerlines after vectorization. A restrained, layout-aware preprocessing pipeline and light augmentation were sufficient to promote robustness, in line with guidance from the George B.\ Moody PhysioNet Challenge 2024 on ECG image digitization.

From a systems perspective, synchronizing the full nnU-Net each round is communication-heavy but simple and effective. The principal gains observed with \emph{FedAdam} required neither partial model exchange nor bespoke compression. When bandwidth is constrained, quantization or sparsification can be layered onto this pipeline, but such measures should be co-tuned with clipping and privacy noise to avoid compounding losses. More broadly, medical-imaging deployments commonly favor cross-silo FL to respect data-locality and governance constraints, and our results reinforce that strong centralized baselines can be approached without pooling data.\cite{Kaissis2020NMI}

\paragraph{Limitations and opportunities.}
First, the privacy analysis reflects \emph{central} DP applied to site-level aggregates. In typical cross-silo settings, a client holds records for many patients; the absence of per-user clipping and accounting at the client guarantees is effectively client-level rather than user-level.\cite{Abadi2016DPSGD,Mironov2017RDP} Second, communication overheads remain nontrivial for full-model synchronization. Future work could include (i) per-user DP accounting within clients' DP-SGD with secure aggregation, (ii) update compression co-designed with clipping and noise, and (iii) personalization and domain generalization to further reduce cross-site dispersion. Finally, evaluation on real scanned ECGs from community benchmarks can broaden external validity and stress-test robustness beyond rendered pages.\cite{Reyna2024Challenge}

\section{Conclusion}
\label{sec:conclusion}
This study evaluates a privacy-aware, cross-silo ECG page–to–waveform digitization framework that trains nnU-Net end-to-end at each site and aggregates full-model updates without sharing images. Across realistic non-IID client splits, adaptive server optimization with \textbf{FedAdam} consistently accelerated learning and achieved the highest late-round plateau (Dice at R100: \emph{FedAdam} 0.935, \emph{FedProx} 0.926, \emph{FedAvg} 0.912), approaching the centralized reference (0.938) while preserving data locality. 

Combining secure aggregation with central Gaussian differential privacy and Rényi accounting maintained competitive accuracy and yielded auditable, deployment-oriented guarantees; the server observed only a clipped, weighted sum of client updates, and calibrated noise was applied post-aggregation. The end-to-end pipeline—layout-preserving normalization, thin-structure segmentation, and calibration-aware vectorization—translated mask continuity gains into steadier twelve-lead reconstructions.

A key limitation concerns the granularity of privacy guarantees in cross-silo settings: clipping and noising a single site-level update provide client-level rather than strict user-level DP unless per-user clipping/accounting is incorporated client-side. Future work will explore per-user DP mechanisms and communication-efficient personalization to further narrow the federated–centralized gap.

\section{Dataset availability}
\label{sec:dataset_availability}
All experiments in this study use the PTB-XL dataset as the authoritative source of twelve-lead ECG waveforms. PTB-XL is a publicly available dataset released under an open license via PhysioNet, with standardized WFDB records and rich metadata suitable for benchmarking and reproducible research \cite{wagner2020ptbxl}. In addition, the 2024 George~B.\ Moody PhysioNet Challenge featured page-image digitization and image-based ECG classification tasks derived from PTB-XL signals, further consolidating PTB-XL as a community reference for image-to-signal reconstruction and downstream modeling. The original PTB-XL waveforms are accessible to the public through PhysioNet. Challenge materials and proceedings provide complementary artifacts and documentation for image rendering and evaluation protocols \cite{physionet2024web}.



\bibliographystyle{ieeetr}

\end{document}